\documentclass[aps,prd,twocolumn,showpacs,superscriptaddress]{revtex4-2}

\usepackage{amsfonts,mathrsfs,amsmath,amsthm,amssymb}
\usepackage{newtxmath}
\usepackage{bm,times}
\usepackage{graphicx,epsfig}
\usepackage[colorlinks=true,breaklinks=true,linkcolor=blue,citecolor=blue,urlcolor=blue]{hyperref}
\usepackage{changes}

\hypersetup{
  pdfauthor={Tao Zhou},
  pdfkeywords={},
  pdftitle={Lorentz gauge and Coulomb gauge for tetrad field of gravity},
  pdfsubject={Research Paper},
  pdfpagemode=UseNone
}

\def \non {\nonumber}
\def \d {\mathrm{d}}

\newcommand{\be}{\begin{eqnarray}}
\newcommand{\ee}{\end{eqnarray}}

\newcommand{\im}{\mathrm{i}}
\newcommand{\udt}[3]{#1^{#2}_{\phantom{#2}#3}}
\newcommand{\dut}[3]{#1_{#2}^{\phantom{#2}#3}}
\newcommand{\dudt}[4]{#1_{#2\phantom{#3}#4}^{\phantom{#2}#3}}

\makeatletter
    
    \newcommand{\Rmnum}[1]{\expandafter\@slowromancap\romannumeral #1@}
\makeatother

\begin{document}
\title{Lorentz gauge and Coulomb gauge for tetrad field of gravity}

\author{Tao Zhou}
\email{taozhou@swjtu.edu.cn}
\affiliation{Quantum Optoelectronics Laboratory, School of Physical Science and Technology, Southwest Jiaotong University, Chengdu 610031, China}
\affiliation{Department of Applied Physics, School of Physical Science and Technology, Southwest Jiaotong University, Chengdu 611756, China}

\date{\today}

\begin{abstract}
In general relativity, an inertial frame can only be established in a small region of spacetime, and locally inertial frames are mathematically represented by a tetrad field in gravity. The tetrad field is not unique due to the freedom to perform Lorentz transformations in locally inertial frames, and there exists freedom to choose the locally inertial frame at each spacetime point. The local Lorentz transformations are known as non-Abelian gauge transformations for the tetrad field, and to fix the gauge freedom, corresponding to the Lorentz gauge $\partial^\mu\mathcal{A}_\mu=0$ and Coulomb gauge $\partial^i\mathcal{A}_i=0$ in electrodynamics, the Lorentz gauge and Coulomb gauge for the tetrad field are proposed in the present work. Moreover, properties of the Lorentz gauge and Coulomb gauge for tetrad field are discussed, which show the similarities to those in electromagnetic field.
\end{abstract}

\pacs{04.20.Cv, 11.15.-q}

\maketitle

\textit{Introduction.---}
Spacetime is flat in special relativity, and a globally inertial frame (GIF) can always be established. Any observer with a constant three-velocity in GIF is an inertial observer, and the laws of physics look the same to all such observers. In general relativity, one main issue, that no unique globally inertial frame can be established, arises from flat to curved spacetime. However, in small regions of spacetime, the spacetime looks approximately flat, and the freely falling observers, who perform physics experiments in these small regions of spacetime around them, would see that physics is approximately the same as that seen by an inertial observer in flat spacetime. Therefore, to retain the notation of inertial frames, one can define \textit{locally inertial frames} (LIFs) in the small enough regions of spacetime, and they cannot be globally extended throughout the spacetime.

The LIFs are described by the tetrad field~\cite{Space_Carroll,Geo_Nakahra}, which will be briefly discussed in the following. The freedom of tetrad field due to the local Lorentz transformations (LLTs) provides different LIFs at the same spacetime point, and the choice of a LIF at the same spacetime point is accordingly quite arbitrary. In the view of gauge theory, the Lorentz transformations play the role of gauge transformations. The redundant gauge freedoms compensate elegant mathematical expressions, but meanwhile, result in the obstacle to identify the real physical degrees of freedom. In electrodynamics, the vector potential $\mathcal{A}^\mu$ is used to describe the massless photon, where the Lorentz gauge $\partial^\mu\mathcal{A}_\mu=0$ and Coulomb gauge $\partial^i\mathcal{A}_i=0$ can both remove certain degrees of the gauge freedom. Lorentz gauge, however, leaves some degrees of unphysical freedom, while Coulomb gauge is sufficient to specify the two physical degrees of photon's  polarization. Similar to that in electrodynamics, the Coulomb gauge for gravity in weak field approximation was proposed and discussed in Refs.~\cite{PhysRevD.83.061501,PhysRevD.83.084006}.

Due to the freedom to the local choice of inertial frame in curved spacetime, there is no way to determine the same direction of the LIFs at two separated spacetime points. This is a dramatic problem, for instance, the spin entanglement state shared by two separated parties has an ambiguous meaning in curved spacetime. To overcome this issue, a gauge condition can be introduced to remove the degrees of freedom as usual in gauge theory, and an appropriate gauge condition leaving only physical freedom is further appreciated. In this paper, corresponding to those in electrodynamics, we investigate the Lorentz gauge and Coulomb gauge for tetrad field, and the properties of the two gauge conditions will be discussed.

\textit{Mathematical representation of LIFs.---}
In general relativity, spacetime is a four-dimensional Lorentzian manifold~$\mathcal{M}$ with a metric tensor $g_{\mu\nu}(x)$, and the tangent space $T_x\mathcal{M}$ at some point $x\in \mathcal{M}$ is spanned by the natural basis $\{e_\mu=\partial/\partial x^\mu\}$. LIFs can be mathematically represented by a tetrad field $\dut{e}{\alpha}{\mu}(x)$~\cite{Space_Carroll,Geo_Nakahra}, which gives a set of orthonormal basis vectors of a LIF, say
\be
e_{\hat{\alpha}}=\dut{e}{\alpha}{\mu}\partial_\mu\ \ \ \ \{\dut{e}{\alpha}{\mu}\}\in\mathrm{GL}(4,\mathbb{R}).
\ee
$\{e_{\hat{\alpha}}\}$ are usually called \textit{non-coordinate bases}, and the orthonormality is satisfied by
\be
\label{orthcon}
g(e_{\hat{\alpha}},e_{\hat{\beta}})=g_{\mu\nu}\dut{e}{\alpha}{\mu}\dut{e}{\beta}{\nu}=\eta_{\alpha\beta},
\ee
where $\eta_{\alpha\beta}$ is Minkowski metric with signature $(-,+,+,+)$. [In this paper, for tetrad field, Greek indices $\alpha$, $\beta$, $\gamma$, $\delta,...$ run over the four spacetime inertial coordinate labels, $\mu$, $\nu$, $\kappa$, $\lambda,...$ run over the four coordinate labels in general coordinate, Latin indices run from $1$ to $3$, and repeated indices are summed over]. As stressed before, the LIF at a point is not unique, and a Lorentz transformation $\Lambda(x)$ can give another tetrad field 
\be
\dut{{e'}}{\alpha}{\mu}=\dut{e}{\beta}{\mu}\udt{{(\Lambda^{-1})}}{\beta}{\alpha},
\ee
and new orthogonal basis
\be
e'_{\hat{\alpha}}=\dut{{e'}}{\alpha}{\mu}\partial_\mu=e_{\hat{\beta}}\udt{{(\Lambda^{-1})}}{\beta}{\alpha}.
\ee
It can be easily checked that the orthonormality of the basis are preserved by the Lorentz transformation, as required.

Actually, a non-coordinate basis at each spacetime point spans a vector space with the Minkowski metric $\eta_{\alpha\beta}$. In the language of differential geometry, the tetrad fields in the spacetime constitute a fiber bundle with Minkowski space a fiber space at each spacetime point, and the Lorentz group $\mathrm{SO}(1,3)$ is the structure group.

\textit{Lorentz connection one-form.---}
A connection $\omega_\mu$ is required to define a covariant derivative in curved spacetime, such as the affine connection in the general coordinate, and the connection coefficients  with respect to the non-coordinate bases $\{e_{\hat{\alpha}}\}$ are defined by
\be
\label{Connec}
\mathcal{D}_\mu e_{\hat{\alpha}}=\dudt{\omega}{\mu}{\beta}{\alpha}e_{\hat{\beta}},
\ee
where $\mathcal{D}_\mu$ is the covariant derivative in LIF. The covariant derivative of a vector $V=V^{\alpha}e_{\hat{\alpha}}$ in LIF is $\mathcal{D}V=\mathcal{D}_\nu V^\alpha\d x^\nu\otimes e_{\hat{\alpha}}$, and according to Eq.~\eqref{Connec}, the components can be easily derived
\be
\mathcal{D}_\mu V^\alpha=\partial_\mu V^\alpha+\dudt{\omega}{\mu}{\alpha}{\beta}V^\beta.
\ee
Now, one can consider the covariant derivative of a vector $V=V^\mu \partial_\mu$ in the general coordinate
\be
\nabla V&=&\nabla_\nu V^\mu \d x^\nu\otimes\partial_\mu\non,
\ee
where the components are
\be
\nabla_\nu V^\mu=\partial_\nu V^\mu+\Gamma^\mu_{\nu\lambda}V^\lambda,
\ee
with $\Gamma^\mu_{\nu\lambda}=\frac{1}{2}g^{\mu\sigma}(\partial_\nu g_{\lambda\sigma}+\partial_\lambda g_{\sigma\nu}-\partial_\sigma g_{\nu\lambda})$ the Levi-Civita connection coefficients, and one can have
\be
\nabla_\nu V^\mu=\dut{e}{\alpha}{\mu}\mathcal{D}_\nu V^\alpha.
\ee 
Since the vector $V$ can also be expressed as $V^\mu=\dut{e}{\alpha}{\mu}V^\alpha$ in LIF, the covariant derivative can be rewritten as
\be
\nabla_\nu V^\mu&=&\nabla_\nu(\dut{e}{\alpha}{\mu}V^\alpha)=\partial_\nu(\dut{e}{\alpha}{\mu}V^\alpha)+\Gamma^\mu_{\nu\lambda}(\dut{e}{\alpha}{\lambda}V^\alpha),\non\\
&=&\dut{e}{\alpha}{\mu}\big[\partial_\nu V^\alpha+(\udt{e}{\alpha}{\lambda}\partial_\nu\dut{e}{\beta}{\lambda}+\dut{e}{\beta}{\lambda}\udt{e}{\alpha}{\sigma}\Gamma^\sigma_{\nu\lambda})V^\beta\big],
\ee
where $\udt{e}{\alpha}{\mu}$ is the inverse of $\dut{e}{\alpha}{\mu}$. Therefore, one can find
\be
\dudt{\omega}{\mu}{\alpha}{\beta}=\udt{e}{\alpha}{\lambda}\partial_\mu\dut{e}{\beta}{\lambda}+\dut{e}{\beta}{\lambda}\udt{e}{\alpha}{\sigma}\Gamma^\sigma_{\mu\lambda},
\ee
which are named the Lorentz (or spin) connection coefficients~\cite{Space_Carroll,Geo_Nakahra}, and $\dut{\omega}{\mu}{\alpha\beta}=-\dut{\omega}{\mu}{\beta\alpha}$. Moreover, the non-coordinate bases $\{e_{\hat{\alpha}}\}$ satisfy the commutation relation $[e_{\hat{\alpha}},e_{\hat{\beta}}]=\udt{f}{\gamma}{\alpha\beta}e_{\hat{\gamma}}$, with $\udt{f}{\gamma}{\alpha\beta}$ the coefficients of the anholonomy of tetrads~\cite{aldrovandi2012teleparallel}, and the spin connection coefficients can also be expressed as~\cite{Geo_Nakahra,aldrovandi2012teleparallel}
\be
\dudt{\omega}{\mu}{\alpha}{\beta}=\frac{1}{2}\left(\dudt{f}{\beta}{\alpha}{\gamma}+\dudt{f}{\gamma}{\alpha}{\beta}-\udt{f}{\alpha}{\beta\gamma}\right)\udt{e}{\gamma}{\mu}.
\ee

Now, one can further define $\omega=\omega_\mu\d x^\mu$ as a Lie-algebra-valued connection one-form, and $\omega_\mu$ take values in the Lie algebra $\mathfrak{so}(1,3)$ of the Lorentz group in the non-coordinate basis given by the tetrad. Denote the algebra generators of $\mathfrak{so}(1,3)$ by $J_{\alpha\beta}$, and then
\be
\omega_\mu\equiv-\frac{\im}{2}\dut{\omega}{\mu}{\alpha\beta}J_{\alpha\beta}.
\ee
In the non-coordinate basis, $J_{\alpha\beta}$ act as the vector representation of the Lorentz generators
\be
\udt{\left(J_{\alpha\beta}\right)}{\gamma}{\delta}=\im\left(\eta_{\beta\delta}\udt{\eta}{\gamma}{\alpha}-\eta_{\alpha\delta}\udt{\eta}{\gamma}{\beta}\right),
\ee
which shows that $\omega_\mu$ here are $4\times4$ matrices with the matrix elements $\udt{(\omega_\mu)}{\alpha}{\beta}=\dudt{\omega}{\mu}{\alpha}{\beta}$. Therefore, $\omega=\omega_\mu\d x^\mu$ is a matrix-valued one-form that can be defined by~\cite{Geo_Nakahra} 
\be
\udt{\omega}{\alpha}{\beta}\equiv\dudt{\omega}{\mu}{\alpha}{\beta}\d x^\mu.
\ee

It is easy to see that $\omega_\mu$ transforms as a vector under a general coordinate transformation $x'^\mu=x'^\mu(x)$,
\be
\label{GCT}
\omega'_\mu(x')=\frac{\partial x^\nu}{\partial x'^\mu}\omega_\nu(x).
\ee 
Under a LLT $e'_\alpha(x)=e_{\hat{\beta}}\udt{{(\Lambda^{-1})}}{\beta}{\alpha}$, however, Lorentz connection transforms inhomogeneously
\be
\label{gaugetrans}
\omega'_\mu(x)=\Lambda(x)\omega_\mu(x)\Lambda^{-1}(x)+\Lambda(x)\partial_\mu\Lambda^{-1}(x).
\ee
For the details of tetrad field and Lorentz connection, see appendix J of Ref.~\cite{Space_Carroll}.

\textit{Lorentz gauge and Coulomb gauge for tetrad fields.---} To obtain the gauge condition for tetrad field, we first briefly review the Lorentz gauge in electrodynamics. Maxwell's theory of electromagnetism is described by the $\mathrm{U}(1)$ gauge group, and the connection one-form, physically representing the photon, is 
\be
\label{cononeform}
\mathcal{A}=\mathcal{A}_\mu \d x^\mu,
\ee
where $\mathcal{A}_\mu$ is the vector potential in electrodynamics. Then, one may naively guess the Lorentz gauge for tetrad field is $\partial^\mu\omega_\mu=0$, corresponding to $\partial^\mu\mathcal{A}_\mu=0$ in electrodynamics, and however, this obviously does not work for tetrad field. For one thing, the base manifold in Maxwell's theory is a flat spacetime, while the one in tetrad field is a curved spacetime, and the ordinary derivative should be replaced by the covariant one. For another, the connection $\omega_\mu$ is an element of Lie algebra $\mathfrak{so}(1,3)$, and the derivative could be further amended to be the gauge-covariant derivative in adjoint representation. To see how to construct the corresponding Lorentz gauge condition, we will further see the Maxwell's theory more mathematically in the following.

With the connection one-form in Eq.~\eqref{cononeform}, the field strength for Maxwell's theory of electromagnetism can be introduced
\be
\mathcal{F}&\equiv&\d\mathcal{A}=\frac{1}{2}\mathcal{F}_{\mu\nu}\d x^\mu\wedge\d x^\nu\non\\
&=&(\partial_\mu\mathcal{A}_\nu-\partial_\nu\mathcal{A}_\mu)\d x^\mu\otimes\d x^\nu,
\ee
and the Bianchi identity
\be
\label{Bianchi}
\d\mathcal{F}=0
\ee
reduces to two of the Maxwell's equations,
\be
\label{Maxwell1}
\bm\partial\times\bm E+\frac{\partial\bm B}{\partial t}=0,\ \ \ \bm\partial\cdot\bm B=0.
\ee
For the other two of Maxwell's equations (with $\varepsilon_0=\mu_0=1$)
\be
\label{Maxwell2}
\bm\partial\times\bm B-\frac{\partial\bm E}{\partial t}=\bm j,\ \ \ \bm\partial\cdot\bm E=\rho,
\ee 
where $\rho$ and $\bm j$ is the electric charge density and electric current density, respectively, to obtain the compact expressions corresponding to Eq.~\eqref{Bianchi}, the adjoint of the exterior derivative (codifferential operator) in the four-dimensional Lorentzian manifold should first be introduced, say
\be
\d^\dag=\ast\d\ast,
\ee
where $\ast$ is the \textit{Hodge operator}~\cite{Geo_Nakahra}. Therefore, $\d^\dag\mathcal{F}=\partial^{\mu}\mathcal{F}_{\mu\nu}\d x^\nu$, and Eq.~(\ref{Maxwell2}) can be straightforwardly rewritten as
\be
\label{Max2}
\d^\dag\mathcal{F}=j,
\ee
with the one-form $j=-\rho\d t+\bm j\cdot\d\bm x$. Furthermore, the Lorentz gauge $\partial^\mu\mathcal{A}_\mu=0$ in electrodynamics now can be expressed as
\be
\label{LG}
\d^\dag\mathcal{A}=0
\ee 
and mathematically, the corresponding Lorentz gauge condition for tetrad field is
\be
\label{Lorentz}
\mathfrak{D}^\dag\omega=0,
\ee
where $\mathfrak{D}=\d+[\omega,\ ]$ is the gauge-covariant derivative in adjoint representation, and $\mathfrak{D}^\dag=\ast\mathfrak{D}\ast$ is the codifferential operator for the four-dimensinal Lorentzian manifold here. To make Eq.~\eqref{Lorentz} more explicit, we first come to the action of $\mathfrak{D}^\dag$ on a Lie-algebra-valued one-form $\mathcal{\varpi}=\varpi_\mu\d x^{\mu}$,
\be
\label{Deri}
\mathfrak{D}^\dag\varpi&=&\ast\mathfrak{D}\ast(\varpi_\mu\d x^\mu)\non\\
&=&\ast\mathfrak{D}\bigg(\frac{\sqrt{g}}{3!}\varpi_\mu g^{\mu\lambda}\varepsilon_{\lambda\nu_2\nu_3\nu_4}\d x^{\nu_2}\wedge\d x^{\nu_3}\wedge\d x^{\nu_4}\bigg)\non\\
&=&\ast\frac{1}{3!}\mathfrak{D}_\nu\big(\sqrt{g}g^{\mu\lambda}\varpi_\mu\big)\varepsilon_{\lambda\nu_2\nu_3\nu_4}\d x^{\nu}\wedge\d x^{\nu_2}\wedge\d x^{\nu_3}\wedge\d x^{\nu_4}\non\\
&=&\ast\mathfrak{D}_\nu\big(\sqrt{g}g^{\mu\nu}\varpi_\mu\big)\d x^0\wedge\d x^1\wedge\d x^2\wedge\d x^3\non\\
&=&\mathfrak{D}_\nu\big(\sqrt{g}g^{\mu\nu}\varpi_\mu\big)\sqrt{g}\varepsilon^{0123}\non\\
&=&-\frac{1}{\sqrt{g}}\mathfrak{D}_\nu\big(\sqrt{g}g^{\mu\nu}\varpi_\mu\big)\non\\
&=&-\frac{1}{\sqrt{g}}\partial_\nu\big(\sqrt{g}g^{\mu\nu}\varpi_\mu\big)-g^{\mu\nu}[\omega_\nu,\varpi_\mu]\non\\
&=&-g^{\mu\nu}\nabla_\mu\varpi_\nu-g^{\mu\nu}[\omega_\mu,\varpi_\nu]
\ee
where $g=|\mathrm{det}\ g_{\mu\nu}|$, $\varepsilon_{\mu_1\mu_2\mu_3\mu_4}$ is the totally anti-symmetric Levi-Civita symbol, and $\varepsilon^{\mu_1\mu_2\mu_3\mu_4}=-g^{-1}\varepsilon_{\mu_1\mu_2\mu_3\mu_4}$ has been used. Now, according to the derivation in Eq.~\eqref{Deri}, the Lorentz gauge in Eq.~\eqref{Lorentz} becomes
\be
\label{Lorentz2}
g^{\mu\nu}\nabla_\mu\omega_\nu=\frac{1}{\sqrt{g}}\partial_\mu\big(\sqrt{g}g^{\mu\nu}\omega_\nu\big)=0.
\ee

Before one comes to the Coulomb gauge for the tetrad field, we first consider the corresponding expression to Lorentz gauge in Eq.~\eqref{LG} for Coulomb gauge. Symbolically, the exterior derivative can be written as
\be
\d\equiv\d t\wedge\frac{\partial}{\partial t}+{\bm \d},
\ee
where ${\bm \d}$ is the \emph{spatial} exterior derivative~\cite{frankel2011geometry}, and the partial differential operator acts only on the coefficients. Since
\be
\d^\dag\mathcal{A}=\partial^0\mathcal{A}_0+\bm\d^\dag\bm{\mathcal{A}},
\ee
with $\bm{\mathcal{A}}=\mathcal{A}_i\d x^i$, $\bm\d^\dag=\ast\bm\d\ast$ and $\bm\d^\dag\bm{\mathcal{A}}=\partial^i\mathcal{A}_i$, the Coulomb gauge in the electrodynamics is
\be
\bm\d^\dag\bm{\mathcal{A}}=0.
\ee

According to the relationship between Lorentz gauge $\partial^\mu\mathcal{A}_\mu=0$ $(\d^\dag\mathcal{A}=0)$ and Coulomb gauge $\partial^i\mathcal{A}_i=0$ $(\bm\d^\dag\bm{\mathcal{A}}=0)$ in electrodynamics, the Coulomb gauge for the tetrad field can be similarly generalized as
\be
\label{Coulomb}
\bm{\mathfrak{D}}^\dag\bm\omega=0,
\ee
with $\bm\omega=\omega_i\d x^i$ and $\bm{\mathfrak{D}}=\bm\d+[\bm\omega,\ ]$, and the gauge-covariant derivative in adjoint representation can be decomposed as
\be
\mathfrak{D}=\d t\wedge\bigg(\frac{\partial}{\partial t}+[\omega_0,\ ]\bigg)+\bm{\mathfrak{D}}.
\ee
By some algebra, it is easy to obtain the explicit expression for the Coulomb gauge in Eq.~\eqref{Coulomb}
\be
\label{Coulomb2}
\frac{1}{\sqrt{g}}\partial_i\big(\sqrt{g}g^{ij}\omega_j\big)=0.
\ee

\textit{Remarks and discussion.---}
(i) Fermi-Walker transport is usually employed to define the LIFs in curved spacetime, and the LIFs are established by parallel transport of the non-coordinate bases in curved spacetime~\cite{misner1973gravitation}. The LIF of an observer in curved spacetime is then generally dependent on the path along which the non-coordinate bases are transported, for instance, the $z$ axes of LIFs at the same spacetime point could be different for the observers come along different path. More seriously, a locally-defined physical object, such as a localized qubit state, at a spacetime can not be defined unambiguously as the LIFs are dependent on the observer's path, and this seems to be ``weird". However, if a gauge condition is carefully chosen, and the gauge freedom of the tetrad fields is fixed, the LIF at each spacetime point can be unique up to a global gauge transformation. The LIFs now are independent on the observer's path, and one has a path-independent way of constructing \textit{``global'' LIFs}. Subsequently, the local physical object can be unambiguously defined with the ``global'' LIFs.

(ii) To determine the LIFs in the curved spacetime, the sixteen functions $\dut{e}{\alpha}{\mu}(x)$ of tetrad field should be fixed. The orthonormality condition in Eq.~\eqref{orthcon} represents ten constrains on the tetrad field, so there are still six degrees of freedom for the tetrad field, which are exactly the three degrees of pure boosts and three degrees of the spatial rotations for the gauge transformation $\Lambda(x)$. As the Lorentz connection $\omega_\mu(x)$ is an antisymmetric matrix, both the Lorentz gauge and Coulomb gauge in Eq.~\eqref{Lorentz2} and Eq.~\eqref{Coulomb2} exactly contain six constrains, and roughly speaking, seem to fix the gauge transformation completely. However, more rigorously, under a gauge transformation in Eq.~\eqref{gaugetrans}, to preserve Lorentz gauge condition in Eq.~\eqref{Lorentz2}, it is required that
\be
g^{\mu\nu}\nabla_\mu(\Lambda\partial_\nu\Lambda^{-1})+g^{\mu\nu}[\Lambda\omega_\mu\Lambda^{-1},\Lambda\partial_\nu\Lambda^{-1}]=0,
\ee
and obviously, the solution for the gauge transformation $\Lambda(x)$ satisfying the equation above is not unique, at least for some simple cases. Therefore, the Lorentz gauge for tetrad field does not fix the gauge completely in general, and the same result holds for Coulomb gauge. Nevertheless, the degrees of freedom unconstrained by the two gauges in tetrad field should have different meanings, but it is not easy to completely clarify the differences due to the complication from Abelian to non-Abelian gauge theory, and only brief discussions about the disparate characteristics of the two gauge conditions are provided in the next.

(iii) In a static case, Lorentz gauge $\partial^\mu\mathcal{A}_\mu=0$ coincides with Coulomb gauge $\partial^i\mathcal{A}_i=0$ in electrodynamics, but for tetrad field in gravity, one does not have the same result. For the spacetime with static metric, Lorentz gauge in Eq.~\eqref{Lorentz2} reduces to $g^{i0}\nabla_i\omega_0-g^{0\mu}\Gamma^\nu_{0\mu}\omega_\nu+g^{ij}\nabla_i\omega_j=0$, and it is different from Coulomb gauge in Eq.~\eqref{Coulomb2} by a term $g^{i0}\nabla_i\omega_0-g^{0\mu}\Gamma^\nu_{0\mu}\omega_\nu$. Moreover, Lorentz gauge for tetrad field is invariant under a general coordinate transformation in Eq.~\eqref{GCT}, while Coulomb gauge is not, and only invariant under a spatial coordinate transformation. These show that the two gauge conditions undoubtedly have distinct properties for tetrad field as well, and one may wonder which is the preferred option in relevant physics problems. Illuminated by the gauge theory of electromagnetism and the proposals in the previous relevant works~\cite{PhysRevD.83.061501,PhysRevD.83.084006,PhysRevLett.100.232002,PhysRevLett.103.062001}, it is reasonable to choose Coulomb gauge as the physical gauge condition. Specifically, the non-coordinate basis $\{e_{\hat{\alpha}}\}$ obeys
\be
\mathcal{D}_\mu\mathcal{D}_\nu e_{\hat{\alpha}}=\udt{{\big[(\nabla_\mu\omega_\nu+\omega_\mu\omega_\nu)\big]}}{\beta}{\alpha}e_{\hat{\beta}}.
\ee
Denote $\bm e=(e_{\hat{0}},e_{\hat{1}},e_{\hat{2}},e_{\hat{3}})$, and in Lorentz gauge,
\be
\label{Leq}
g^{\mu\nu}\mathcal{D}_\mu\mathcal{D}_\nu \bm e-\bm eg^{\mu\nu}\omega_\mu\omega_\nu=0,
\ee
while in Coulomb gauge, one has
\be
\label{Ceq}
g^{ij}\mathcal{D}_i\mathcal{D}_j \bm e-\bm e g^{ij}\omega_i\omega_j=0.
\ee
Comparing Eq.~\eqref{Ceq} with Eq.~\eqref{Leq}, it shows that, for Coulomb gauge, the non-coordinate basis $\{e_{\hat{\alpha}}\}$ is non-dynamic and instantaneously does not propagate in spacetime, which means the inertial effects from the freedom of tetrad field are absent, leaving only the gravitational effects on the LIFs. As a result, established by the tetrad field under Coulomb gauge, the LIF stands a good chance to have advantages over the other ones, and this will be our further investigation elsewhere.

(iv) In gauge theory, the decomposition of gauge potential plays an important role in both theoretical physics and mathematics~\cite{Sci.Sin.11.1072,Sci.Sin.1.45}, and recently, a gauge decomposition approach was proposed to find a gauge-covariant description of the gluon spin and orbital angular momentum in Ref.~\cite{PhysRevLett.100.232002}, and further developed in Ref.~\cite{PhysRevLett.103.062001}. It is a familiar practice to decompose the Lorentz connection into the physical part and pure-gauge part. Mathematically, one can have the separation $\omega_\mu(x)=\hat{\omega}_\mu(x)+\bar{\omega}_\mu(x)$, with $\hat{\omega}_\mu(x)$ the physical part while $\bar{\omega}_\mu(x)$ the pure-gauge part, and under the gauge transformation in Eq.~\eqref{gaugetrans},
\begin{subequations}
\be
\hat{\omega}'_\mu(x)&=&\Lambda(x)\hat{\omega}_\mu(x)\Lambda^{-1}(x),\\
\bar{\omega}'_\mu(x)&=&\Lambda(x)\bar{\omega}_\mu(x)\Lambda^{-1}(x)+\Lambda(x)\partial_\mu\Lambda^{-1}(x).
\ee
\end{subequations}
The equations to define the decomposition can be analogously constructed as
\begin{subequations}
\be
\frac{1}{\sqrt{g}}\partial_i\big(\sqrt{g}g^{ij}\hat{\omega}_j\big)+g^{ij}[\bar{\omega}_i,\hat{\omega}_j]&=&0,\\
\partial_\mu\bar{\omega}_\nu-\partial_\nu\bar{\omega}_\mu+[\bar{\omega}_\mu,\bar{\omega}_\nu]&=&0.
\ee
\end{subequations}

(v) Parallelly to Maxwell's equations~\eqref{Maxwell1} and~\eqref{Maxwell2} in electrodynamics, one may wonder analogical equations for tetrad field in gravity. We would first come to the field strength of tetrad field
\be
\label{Rstrength}
\mathcal{R}\equiv\d\omega+\omega\wedge\omega=\frac{1}{2}\mathcal{R}_{\mu\nu}\d x^\mu\wedge\d x^\nu,
\ee
where $\mathcal{R}_{\mu\nu}$ are the Lie-algebra-valued components of the field strength
\be
\mathcal{R}_{\mu\nu}=\partial_\mu\omega_\nu-\partial_\nu\omega_\mu+[\omega_\mu,\omega_\nu],
\ee
and we can also have the curvature two-form $\udt{\mathcal{R}}{\alpha}{\beta}=\frac{1}{2}\udt{R}{\alpha}{\beta\mu\nu}\d x^\mu\wedge\d x^\nu$. With Eq.~{\eqref{Rstrength}}, the Bianchi identity again gives $\mathfrak{D}\mathcal{R}=0$, reducing to
\be
\d\mathcal{R}+\omega\wedge\mathcal{R}-\mathcal{R}\wedge\omega=0,
\ee
which is the analogical equation to Eq.~\eqref{Maxwell1}. It is a little more complicated to reach the similar results in Eq.~\eqref{Maxwell2}, and before that, one may homoplastically have
\be
\mathfrak{D}^\dag\mathcal{R}=\big(\nabla^\mu\mathcal{R}_{\mu\nu}+[\omega^\mu,\mathcal{R}_{\mu\nu}]\big)\d x^\nu.
\ee
Further consider the Einstein's equation
\be
\udt{R}{\alpha}{\mu}=8\pi G(\udt{T}{\alpha}{\mu}-\frac{1}{2}T\udt{e}{\alpha}{\mu}),
\ee
with $\udt{R}{\alpha}{\mu}=\udt{R}{\alpha\beta}{\mu\nu}\dut{e}{\beta}{\nu}$ the Ricci tensor, $T=g^{\mu\nu}T_{\mu\nu}$ the trace of energy-momentum tensor $\udt{T}{\mu}{\nu}$, and $G$ Newton's constant of gravitation, and one can obtain 
\be
\mathfrak{D}^\dag\mathcal{R}=-4\pi G\d\mathcal{T},
\ee
or, more explicitly
\be
\nabla^\mu\mathcal{R}_{\mu\nu}+[\omega^\mu,\mathcal{R}_{\mu\nu}]=-4\pi G\partial_\nu\mathcal{T},
\ee
with the matrix elements $\udt{\mathcal{T}}{\alpha}{\beta}=T\udt{\delta}{\alpha}{\beta}$. This plays the role of dynamic equation for Lorentz connection $\omega_\mu(x)$ of tetrad field. The influence of the Lorentz gauge in Eq.~\eqref{Lorentz2} and Coulomb gauge in Eq.~\eqref{Coulomb2} on the dynamics of the Lorentz connection $\omega_\mu$ is an interesting and important question, and deserves systematically considerations in the future.

\textit{Conclusions and summaries.---}
The gauge aspect of tetrad field in gravity is investigated in this work, and Lorentz gauge and Coulomb gauge are proposed and discussed in the non-Abelian gauge theory of tetrad field. Though the two gauge conditions are much more complicated than the corresponding ones in the $\mathrm{U}(1)$ gauge theory, some analogous features compared with electromagnetic field are revealed in our discussions. Coulomb gauge is more likely to be the physical gauge, and it is preferable to fix the tetrad field. Completion of the gauge theory of tetrad field in gravity is a sizable task that require a thorough consideration. Some related problems are still worth investigations. The applications of the gauge condition in physics are of particular interest and significance, and the details are our further considerations.

\textit{Acknowledgements.---}
This work was supported by the National Natural Science Foundation of China (Grants No.~12147208 and No.~11405136), and the Fundamental Research Funds for the Central Universities (Grant No. 2682021ZTPY050).

\bibliography{refs}

\end{document}